\documentclass[12pt]{article}
\usepackage{amsmath}
\usepackage{amssymb}

\newcommand{\be}{\begin{equation}}
\newcommand{\bea}{\begin{eqnarray}}
\newcommand{\ee}{\end{equation}}
\newcommand{\eea}{\end{eqnarray}}
\def\theequation{\arabic{section}.\arabic{equation}}
\newcommand{\bn}[2]{\left(\begin{array}{c} #1\\ #2
\end{array}\right)}

\begin{document}
\topmargin -1cm \oddsidemargin=0.25cm\evensidemargin=0.25cm
\setcounter{page}0
\renewcommand{\thefootnote}{\fnsymbol{footnote}}
\begin{titlepage}
\vskip .7in
\begin{center}
{\Large \bf  On Tensorial Spaces and  BCFW  Recursion Relations for Higher Spin Fields
 } \vskip .7in {\large  Mirian
Tsulaia} \vskip .2in { \it
Center of Elementary Particle Physics, Institute for Theoretical Physics,
Ilia State University,
Tbilisi 0162, Georgia} \\
%\vspace{.7in}
\begin{abstract}
%\vskip .5in \noindent

In this short review we briefly consider two topics in the higher spin gauge theory:
the method of ``tensorial (super) spaces" and  application of BCFW recursion relations to higher spin fields.

\end{abstract}

\end{center}

\vfill

\end{titlepage}

\tableofcontents

\section{Introduction}

Since  it has been shown \cite{Fradkin:1986qy}--\cite{Vasiliev:1990en} that higher spin gauge theory is a
consistent theory of interacting fields with infinitely growing spins,
 (see \cite{Vasiliev:1999ba}--\cite{Sagnotti:2011qp} for reviews of different aspects of the subject)
this theory is attracting considerable and growing  attention.

In the present review we briefly describe two topics in the higher spin gauge theory.
In the first part of the review we consider
a description of massless higher spin fields in terms of so called ``tensorial"
(super)spaces. This approach assumes an extension of the original  flat or anti de Sitter
spaces in terms of extra coordinates. These extra coordinates correspond to spin degrees of freedom.
The free field dynamics in this extended space is described in terms of a simple wave equation,
which makes the conformal symmetries of the system manifest.

 The idea of introducing extra coordinates for the description
of spin degrees freedom was first expressed in \cite{Fronsdal:1985pd},
where it has been conjectured that $OSp(1|8)$ group
is a symmetry group of full interacting theory of massless higher
spin fields in four dimensions.
The first explicit realization of this idea was given
in \cite{Bandos:1998vz}--\cite{Bandos:1999qf} where a model of
twistorial superparticle on an extended superspace was considered.
In this consideration extra bosonic coordinates were cannonically conjugated to
central charges in the corresponding $SUSY$ algebra.
The quantization  has revealed
that the helicity constraint, which is present in an ordinary superparticle model
and restricts a superspin of the superparticle to a particular value,
 does not exist for a tensorial superparticle model. Thus, the spectrum contains an infinite number of higher spin fields.
In \cite{Vasiliev:2001zy} M. Vasiliev  extensively developed this subject,
 having shown that the first--quantized field equations in
tensorial superspace of a bosonic dimension $n(2n+1)$ and of a
fermionic dimension $2nN$ are $OSp(N|4n)$ invariant, and for $n=2$
correspond to so called unfolded higher spin field equations in
$D=4$. It has also been shown \cite{Vasiliev:2001dc} that the theory
possesses properties of causality and locality.
A detailed analysis of free equations in $D=3,4,6$ and $10$
was further given in \cite{Bandos:2005mb}.
One can also consider a tensorial extension of anti de Sitter spaces
\cite{Didenko:2003aa}--\cite{Plyushchay:2003tj} and in this way obtain
the free unfolded formulation of massless higher spin fields on anti de Sitter
space--time.
Moreover tensorial superspaces are not only an elegant tool for the description
of dynamics of free massless higher spin  fields on flat and anti de Sitter backgrounds.
In  \cite{Gelfond:2010pm} (see also \cite{Vasiliev:2007yc}--\cite{Gelfond:2003vh}) it has been shown how to describe interactions
of higher spin fields with external higher spin currents using this approach.

In the second part of the review we  address a problem of
consistency for the higher spin theory on the flat space--time background.

An interaction of massless higher spin fields on a flat background, unlike the one on AdS space which naturally bypasses
Coleman--Mandula no go theorem is considered to be more problematic. However,
a problem of constructing  cubic interaction vertices
for massless higher spin fields has been extensively studied on the flat space--time background
\cite{Bengtsson:1987jt}--\cite{Sagnotti:2010at} and interesting solutions for cubic vertices have been found.
An important question, however, is if one can build a consistent perturbation theory
for massless higher spin fields on Minkowski space-time in order to obtain a nontrivial
S--matrix, since
 the most formidable problems in the higher spin gauge theory on flat background start from four point interactions.
One can achieve   consistent cubic interactions and  closure of the algebra of gauge transformations (even using a finite number of fields)
at the level ${\it parameter}$ $\times$ $\it{ (field)}^2$.   When going to higher order
interaction vertices one encounters a requirement to introduce of an infinite number
of gauge fields.
Moreover, one needs either to introduce quartic interaction vertices or impose
 some extra conditions on the cubic vertices in order  to close the algebra  at the level $\it{parameter}$ $\times$
$\it{(field)^3}$. These requirements can in turn lead to a trivial  $S$--matrix.
Then one needs to continue this procedure to higher (quintic) levels, and so on.

It would be desirable to use the extensive knowledge of cubic vertices on flat space \cite{Bengtsson:1987jt}--\cite{Sagnotti:2010at}
to obtain  information about the consistency  of higher order interactions
for massless higher spin fields.
For this reason one can follow the strategy given in  \cite{Fotopoulos:2010ay}: use the solutions for cubic interaction vertices given
in \cite{Bengtsson:1987jt},\cite{Sagnotti:2010at} and apply  BCFW
\cite{Britto:2004ap}--\cite{Benincasa:2007xk} recursion relations to them. This way one can express
four point functions in terms of
three point ones, in order to answer the question about the properties (existence) of the corresponding S--matrix.
 A study of four point vertices and
BCFW relations for massless higher spin fields  carried out in \cite{Fotopoulos:2010ay} indicated
 that  consistency of the theory requires an inclusion of  extended, and possibly nonlocal, objects\footnote{See also
 \cite{Sagnotti:2010at}, \cite{Polyakov:2010sk},
 \cite{Taronna:2011kt}
 for studies of four point functions and the discussion of nonlocalities in the theory
 and \cite{Bengtsson:2006pw} for a general strategy of constructing of higher order vertices.}
 like stringy Pomerons \cite{Brower:2006ea} which were used in \cite{Cheung:2010vn}--\cite{Boels:2010bv} to prove BCFW relations for string theory.

The  paper is organized as follows:

In  Section 2 we review the main properties of tensorial spaces (see \cite{Fronsdal:1985pd}--\cite{Gelfond:2010pm}, \cite{Bandos:2004nn}--\cite{Ivanov:2005ss} for more details\footnote{See also \cite{Bastianelli:2008nm} for some
mechanical models which lead to higher spin fields}). We consider a tensorial extension of both
 flat and anti de Sitter spaces and briefly discuss the interaction with higher spin currents in this approach.

 In  Section 3 we briefly review some cubic interaction vertices for massless higher spin fields on the Minkowski background.
 Then, after a short review of BCFW recursion relations, we
  apply BCFW recursion relations to  theories with cubic interaction vertices described earlier.

\setcounter{equation}0\section{Tensorial space}
\subsection{Flat space-time}

Let us formulate the basic idea behind the introduction of tensorial space.
We shall mainly concentrate on a tensorial extension of four dimensional $D=4$
Minkowski space--time. A generalization for higher dimensional $D=6$ and
$D=10$ spaces will be given later in this Section.
Our discussion will include only bosonic tensorial spaces, without  their
supesymmetric generalizations.

Consider a four dimensional massless scalar field. Its light --like momentum
(see for example \cite{Cachazo:2005ga} for a review)
$p_m p^m=0$, $m=0,1,2,3$ can be parametrized using two commuting mutually complex conjugate Weyl spinors
$\lambda_A$ and $\overline \lambda_{\dot A}$ with $A, {\dot A}=1,2$
\begin{equation} \label{BW}
p^m = \lambda^A (\sigma^m)_{A\dot A}\tilde\lambda^{\dot A},  \quad or \quad
    P_{A \dot A} = \lambda_A  \overline \lambda_{\dot A}.
\end{equation}
Obviously since the spinors are commuting, one has $\lambda^A \lambda_A=\lambda^{\dot A} \lambda_{\dot A}=0 $
and therefore $P^{A \dot A}P_{A \dot A}=0$.
In order to generalize these construction to higher dimensions note that
one can equivalently rewrite the equation (\ref{BW}) in terms of
four dimensional real Majorana spinors $\lambda^{\alpha}$, where $\alpha = 1,...,4$
as
\begin{equation} \label{BM}
p^m = \lambda^\alpha  \gamma_{\alpha \beta}^m \lambda^{\beta}
\end{equation}
since due to Fierz identities for Dirac $(\gamma^m)_{\alpha \beta}=(\gamma^m)_{ \beta \alpha}$
matrices
\begin{equation} \label{D101}
(\gamma^m)_{\alpha \beta} (\gamma_m)_{\gamma \delta}+ (\gamma^m)_{\alpha \delta} (\gamma_m)_{\beta \gamma}+
(\gamma^m)_{\alpha \gamma} (\gamma_m)_{\delta \beta} =0\,
\end{equation}
one has $p^m p_m=0$. Let us note that since the identity
(\ref{D101}) is valid
not only in four dimensions but
in $D=3, 4, 6, 10$, the parametrization (\ref{BM})
of a light--like momentum  via commuting
spinors is valid in these dimensions
as well (see \cite{Bandos:1999qf} for a more detailed discussion on this point).

Let us continue with the four dimensional case. The four momentum
$P_{A \dot A} $ is cannonically conjugate to coordinates
$x^{A \dot A}$. One can easily solve the quantum analogue
of the equation (\ref{BW})
\begin{equation} \label{PWE}
 (\frac{\partial}{\partial x^{A \dot A}} - i\lambda_A  \overline \lambda_{\dot A}) \Phi(x, \lambda)=0
\end{equation}
to obtain a
plane wave solution for the massless scalar particle
\begin{equation}\label{PWWS}
\Phi(x, \lambda)= \phi(\lambda) e^{i x^{A \dot A} \lambda_A  \overline \lambda_{\dot A}},
\end{equation}
or in terms of Majorana spinors
\begin{equation}\label{PWMS}
\Phi(x, \lambda)= \phi(\lambda) e^{i x_{m} \lambda^\alpha \gamma^m_{\alpha \beta}  \ \lambda^{\beta}},
\end{equation}
with some unknown function $\phi(\lambda)$.

Let us now consider the equation
\begin{equation}\label{PRE}
P_{\alpha \beta} = \lambda_{\alpha} \lambda_\beta,
\end{equation}
which looks like a straightforward generalization of (\ref{BW}) and see its implications.
A space--time  described by the coordinates $X^{\alpha \beta}$ (conjugate to $P_{\alpha \beta}$)
is now ten dimensional since it is described by a $4 \times 4$ symmetrical matrix.
 A corresponding basis of symmetrical matrices is represented
by four Dirac matrices $\gamma^m_{\alpha \beta}$ and six antisymmetric (in $m$ and $n$)
combinations $\gamma^{mn}_{\alpha \beta}$.
Therefore a general expression
 for the coordinate $X^{\alpha \beta}$ is
 \begin{equation} \label{tensorial}
 X^{\alpha \beta} = \frac{1}{2} x^m (\gamma_m)^{\alpha \beta} + \frac{1}{4} y^{mn}(\gamma_{mn})^{\alpha \beta}.
 \end{equation}
  The analogue of the wave equation (\ref{PWE})
 is now
 \begin{equation} \label{PRR}
( \frac{\partial}{\partial x^{\alpha \beta}} - i\lambda_\alpha  \lambda_{\beta}) \Phi(X, \lambda)=0,
\end{equation}
whereas the solution of the wave equation (\ref{PRR}) has the form
\begin{equation}\label{soll}
\Phi(X,\lambda)=e^{iX^{\alpha \beta}\lambda_\alpha \lambda_\beta}\phi(\lambda).
\end{equation}

 At this point one might ask the question: what is the meaning of the equation  (\ref{PRR}) and of the extra coordinates
 $y^{mn}$? The answer to this question is the following: the equation  (\ref{PRR}) is nothing else but
 Vasiliev's unfolded equations for massless free higher spin fields on four dimensional Minkowski space--time.
 The wave function $\Phi(X, \lambda)$
depends on both coordinates $x^m$ and $y^{mn}$. While $x^m$ describes a customary four dimensional Minkowski space time,
an expansion of the wave function in terms of powers of $y^{mn}$ generates fields with higher spins
on the four dimensional Minkowski space.

In order to demonstrate these statements let us first Fourier transform the wave function (\ref{soll})
into a different representation
\begin{equation}\label{yr}
C(X,Y)=\int\,d^4\lambda \,e^{-iY^\alpha \lambda_\alpha}\Phi(X,\lambda)=\int\,d^4\lambda \,e^{-iY^\alpha \lambda_\alpha+
i X^{\alpha \beta}\lambda_\alpha \lambda_\beta}\phi(\lambda).
\end{equation}
The function $C(X,Y)$ obeys the equation
\begin{equation}\label{Y}
\left({\partial\over{\partial
X^{\alpha \beta}}}-i{\partial^2\over{\partial Y^\alpha \partial
Y^\beta}}\right)C(X,Y)=0.
\end{equation}
Further, in order to make the connection with Vasiliev's equations in $D=4$ let us rewrite
(\ref{PRE}) in Weyl notations
\begin{eqnarray}
P_{AB}=\lambda_A\lambda_B\,,\quad \overline{P}_{\dot{A}\dot{B}}=
\overline{\lambda}_{\dot A}\overline{\lambda}_{\dot B}\,,
\quad P_{A\dot{A}}=\lambda^{}_A\overline{\lambda}_{\dot A}\,,
\end{eqnarray}
therefore
\begin{eqnarray}\label{Yy}
\left(\sigma^{mn}_{AB}{\partial\over{\partial
y^{mn}}}+i{\partial^2\over{\partial Y^A\partial
Y^B}}\right)C(x,y,Y)=0,\,\nonumber\\
\\
\left(\overline{\sigma}^{mn}_{{\dot A}{\dot
B}}{\partial\over{\partial y^{mn}}}-i{\partial^2\over{\partial
\overline{Y}^{\dot A}\partial \overline{Y}^{\dot
B}}}\right)C(x,y,Y)=0\nonumber
\end{eqnarray}
and
\begin{eqnarray}\left(\sigma^m_{A\dot{A}}{\partial\over{\partial
x^m}}+i{\partial^2\over{\partial Y^A\partial \bar{Y}^{\dot
A}}}\right)C(x,y,Y)=0\,.\label{unfold}
\end{eqnarray}
Equations (\ref{Yy}) relate the dependence of $C(x,y,Y)$ on the coordinates
$y^{mn}$ to its dependence on $Y^\alpha$ and using this relation one can
regard the wave function
$C(x^m,Y^\alpha):=C(X^{\alpha \beta},Y^\alpha)|_{y^{mn}=0}$ at $y^{mn}=0$ as the
fundamental field.

The expansion of $C(x^m,Y)$ in terms of $Y^A$ and
$\overline{Y}^{\dot A}$
is
\begin{equation}\label{gener}
C(x^p,Y^A,\overline{Y}^{\dot
A})=\sum_{m,n=0}^{\infty}\frac{1}{m!n!}\, C_{A_1 \ldots
A_m,\,{\dot B}_1 \ldots {\dot B}_n }(x^p)\, Y^{A_1} \ldots Y^{A_m}
\,\overline{Y}^{{\dot B}_1} \ldots \overline{Y}^{{\dot B}_n}\,,
\end{equation}
where  reality  imposes $(C_{A_1 \ldots A_m,\,{\dot B}_1 \ldots
{\dot B}_n })^*=C_{B_1 \ldots B_n,\,{\dot A}_1 \ldots {\dot A}_m }$,
and the spin--tensors $C$ are by definition symmetric in the indices
$A_i$ and in
$\dot B_i$.

The consistency of (\ref{unfold}) implies the integrability
conditions
\begin{eqnarray}
\label{masseq}\frac{\partial^2}{\partial Y^{[A}
\partial
x^{B]\dot B}}\, C(x^{C{\dot C}},Y) =0, \quad
\frac{\partial^2}{\partial \bar{Y}^{[\dot A} \partial x^{{\dot
B}]B}}\, C(x^{C\dot C},Y) =0\,.
\end{eqnarray}

Let us recall   Vasiliev's unfolded formulation of free higher spin fields in terms of zero--forms.
In this formulation the
 $C_{0,0}$ component (a physical scalar),
 $C_{A_1 \ldots A_{2s},0}$ and $C_{0, {\dot A}_1, \ldots {\dot A}_{2s}}$ components of the expansion
(\ref{gener}) correspond to the physical fields, while the other fields are auxiliary.
The latter  two fields are
 the self--dual and anti--self--dual
components  of the spin--$s$ field strength. The
nontrivial equations on the dynamical fields are \cite{Vasiliev:1999ba} the
Klein--Gordon equation for the spin zero scalar field  $\partial^m \partial_m
C_{0,0}=0$  and the massless  equations
for spin $s>0$ field strengths
\begin{equation}\partial^{B\dot B}C_{BA_1\ldots A_{2s-1}}(x)=0\,,\quad
\partial^{B\dot B}C_{{\dot B}{\dot A}_1\ldots {\dot A}_{2s-1}}(x)=0\,,\label{Bw}
\end{equation}
which follow from (\ref{masseq}) \footnote{The well known counting
of the degrees of freedom is as follows: the symmetric tensor
$C_{BA_1\ldots A_{2s-1}}$ has ${2s+1\choose 2s}=2s+1$ components
satisfying  ${{2s}\choose{2s-1}}=2s$ independent conditions; this
leaves in $C_{BA_1\ldots A_{2s-1}}$ one independent helicity
degree of freedom, as is well known for the massless spin $s$
fields in $D=4$. The spin $s$ state of opposite helicity is
described by $C_{{\dot B\dot A}_1\ldots {\dot A}_{2s-1}}(x^m)$.}.

All the components of $C(x^m,Y^A,\overline{Y}^{\dot
A})$ that depend on {\it both} $Y^A$ and $\overline{Y}^{\dot A}$
are auxiliary fields expressed by (\ref{unfold}) in terms of
space--time derivatives of the dynamical fields contained in the
analytic fields $C(x^m,Y^A,0)$ and $C(x^m,0,Y^{\dot A})$
and thus one arrives at the unfolded formulation of
\cite{Vasiliev:1999ba}.

Let us briefly mention an alternative way to derive the same results. This approach  is more convenient
for  further  generalization for $D=6$ and $D=10$. We shall outline the main idea
  and refer to \cite{Bandos:2005mb} for the details.

  Let us rewrite the equation (\ref{gener}) in terms of Majorana
  spinors\footnote{ As we mentioned before one can of course use  either Mayorana or
  Weyl spinors. The equations written in terms of Majorana spinors are simpler to generalise to higher dimensions, while the equations
  written in terms of Weyl spinors are simpler to compare with original four dimensional equations of \cite{Vasiliev:1999ba}.}
  \begin{equation}\label{pol}
C(X,Y)=\sum^\infty_{n=0}C_{\alpha_1\cdots\alpha_n}(X)\,Y^{\alpha_1}\cdots
Y^{\alpha_n}=b(X)+f_\alpha (X)Y^\alpha +\cdots\,.
\end{equation}
and put it into the equation (\ref{Y}). Then one finds that all components of  $C(X,Y)$ proportional to the higher powers
of $Y^\alpha$ are expressed in terms of two fields: $b(X)$ and $f_\alpha(X)$. As a result of (\ref{pol}) these fields satisfy the relations
\cite{Vasiliev:2001zy}
\begin{eqnarray}\label{b}
\partial_{\alpha \beta}
\partial_{\gamma \delta}\,b(X)-\partial_{\alpha \gamma}\partial_{\beta \delta}\,b(X)&=&0\,,\\
\quad \partial_{\alpha \beta} f_\gamma(X)-\partial_{\alpha \gamma}
f_\beta(X)&=&0\label{f}\,.
\end{eqnarray}
Now the basic objects $b(X)$ and $f_\alpha(X)$ depend on both $x^m$ and $y^{mn}$. Let us expand these fields
in terms of the tensorial coordinates
\begin{eqnarray}
b(x,\,y)&=\phi(x)+y^{m_1n_1}F_{m_1n_1}(x)
+y^{m_1n_1}\,y^{m_2n_2}\,{\hat R}_{m_1n_1,m_2n_2}(x)\nonumber\\
&+\sum_{s=3}^{\infty}\,y^{m_1n_1}\cdots y^{m_sn_s}\,{\hat
R}_{m_1n_1,\cdots,m_sn_s}(x)\,,\label{is}
\end{eqnarray}
\begin{eqnarray}
f^\alpha (x,y)&=&\psi^\alpha(x)+y^{m_1n_1}\,{\hat{\cal R}}^\alpha_{m_1n_1}(x )\nonumber \\
&&+\sum_{s={5\over 2}}^{\infty}\,y^{m_1n_1}\cdots y^{m_{s-{1\over
2}} n_{s-{1\over 2}}}\,{\hat{\cal
R}}^\alpha_{m_1n_1,\cdots,m_{s-{1\over 2}}n_{s-{1\over 2}}}(x )\,.
\label{his}
\end{eqnarray}
Each  component field in this expansion is antisymmetric under permutation of indexes $m_i$ and $n_i$, but is symmetrical
with respect to permutation of pairs $(m_i, n_i)$ with $(m_j,n_j)$.
In order to answer the question
about the physical meaning of these fields one can use $\gamma$ - matrix identities in four dimensions
 to rewrite equation (\ref{b})
in an equivalent form
 $$
\partial_p\,\partial^p\,b(x^l,y^{mn})=0, \quad
\left(\partial_p\,\partial_q
-4\,\partial_{pr}\,\partial^r_{~q}\right)\,b(x^l,y^{mn})=0, \quad
\epsilon^{pqrt}\partial_{pq}\,\partial_{rs}\,b(x^l,y^{mn})=0,
$$
\begin{equation} \label{KG}
\epsilon^{pqrt}\partial_q\,\partial_{rt}\,\,b(x^l,y^{mn})=0,
\quad
\partial^{~p}_{q}\,\partial_p\,\,b(x^l,y^{mn})=0\,.
\end{equation}
where $\partial_p = \frac{\partial}{\partial x^p}$ and $\partial_{pq}= \frac{\partial}{\partial y^{pq}}$.
The meaning of the equations (\ref{KG}) is the following:
the first equation is a Klein-Gordon equation. The second equation implies that
a trace of the tensor which  comes with  the $s$- th power of $y^{mn}$
in the expansion (\ref{b})
is expressed via the second derivative of the tensor which comes with
the $s-2$- th power of $y^{mn}$ in the expansion (\ref{b}). Therefore traces are not
independent degrees of freedom and
 the tensorial fields under consideration are effectively traceless.
The third and  fourth equations in (\ref{KG})
mean  that the tensor fields   satisfy the Bianchi identities,
 and the last equation implies  that
they are  co--closed.
These are equations for conformal higher spin fields written in terms of curvatures
${\hat{\cal
R}}^\alpha_{m_1n_1,\cdots,m_{s-{1\over 2}}n_{s-{1\over 2}}}(x )$.
Since in four dimensions equations for higher spin fields are conformally invariant
we obtain field equations for all higher spin fields in $D=4$.
The treatment of the equation (\ref{f}) which describes half integer higher spin fields
in terms of corresponding curvatures is completely analogous to the bosonic one (\ref{b}).

Let us summarize our discussion: in order to describe the dynamics of higher spin fields in four dimensions
we have introduced extended ten dimensional space - so called ``tensorial space" parametrized by coordinates
$X^{\alpha \beta}$ (\ref{tensorial}). The main object is a generating functional for higher spin fields
described by $C(X, Y)$ or by $\Phi(X, \lambda)$. the generating functional depends on tensorial coordinates
$X^{\alpha \beta}$ and on the commuting spinors $Y^\alpha$ or $\lambda^\alpha$.
The dynamic is encoded into field equations (\ref{PRR}) or (\ref{Y}).
To obtain ``actual" field equations in terms of only physical coordinates $x^m$
one can use two different approaches. In the framework of the first approach one first gets rid of
tensorial coordinates $y^{mn}$ and arrives at Vasiliev's unfolded formulation
in terms of the functional (\ref{gener}). Alternatively, one can first get rid of
the commuting spinor variables and arrive at geometric equations
formulated in terms of (\ref{is}) and (\ref{his}).
In both pictures the equations are formulated in terms of field strength
of four dimensional higher spin fields; the difference is that they are represented via either spinorial
or vectorial indexes.

Finally, let us give a definition of tensorial spaces in $D=6$ and $D=10$ space--time
dimensions.
 In $D=10$
 the twistor--like variable $\lambda_\alpha$ is a 16--component
Majorana--Weyl spinor. The gamma--matrices
$\gamma_m^{\alpha \beta}$ and $\gamma_{m_1\cdots m_5}^{\alpha \beta}$ form a basis
of the symmetric $16\times 16$ matrices, so the $n=16$ tensorial
manifold is parametrized by the coordinates
\begin{equation}
X^{\alpha \beta}= {1\over 16}\,\Big(\,x^m\gamma_m^{\alpha \beta}+{1\over {2\cdot
5!}} \,y^{m_1\ldots m_5}\gamma_{m_1\ldots
m_5}^{\alpha \beta}\Big)=X^{\beta \alpha}\,,
\end{equation}
$$
\quad (m=0,1,\ldots,9\,; \quad
\alpha,\beta=1,2,\ldots,16)\,,
$$
where $x^m=X^{\alpha \beta}\gamma^m_{\alpha \beta}\,$ are associated with the
coordinates of the $D=10$ space--time, while the anti--self--dual
coordinates
$$y^{m_1\ldots m_5}=X^{\alpha \beta}\gamma^{m_1\ldots m_5}_{\alpha \beta}=-{1
\over 5!}\,\epsilon^{m_1\ldots m_5n_1\ldots n_5}y_{n_1\ldots
n_5}\,,$$ describe spin degrees of freedom.

The corresponding field equations are again (\ref{b}) and (\ref{f}) and the entire discussion repeats as in the case
of $D=4$. The crucial difference is that now the expansion (\ref{is}) and (\ref{his})
is performed in terms of the coordinates $y^{m_1\ldots m_5}$. As a result
one obtains a description of conformal fields whose curvatures are self--dual
with respect to each set of indexes $(m_1 n_i p_i q_i r_i)$.
These traceless rank $5s$ tensors ${ R}_{[5]_1 \cdots [5]_{s} }$  are
automatically irreducible under $GL(10,\mathbb R)$ due to the
self--duality property, and are thus associated with the
rectangular Young diagrams $(s,s,s,s,s)$ which are made of five rows
of equal length $s$ (``multi --five--forms").
 The field equations, which are ten--dimensional analogues of
four dimensional equations (\ref{KG}), can be found in \cite{Bandos:2005mb}.

In $D =6$ the commuting spinor $\lambda_\alpha$ is  a symplectic Majorana--Weyl
spinor.  The spinor index
can thus be decomposed as
$\alpha=a\otimes i$ ($\alpha=1,\ldots,8$; $a=1,2,3,4$; $i=1,2$). The
tensorial space coordinates $X^{\alpha \beta}=X^{ai\,bj}$ are decomposed
into
\begin{eqnarray}\label{6} && X^{ai\, bj}\, =  \,{1\over 8} \, x^m\, {\tilde\gamma}_m^{a
b} \,\epsilon^{ij}\, + {1\over 16 \cdot 3!}\,y_I^{mnp}\,
{\tilde\gamma}^{ab}_{mnp}\,\tau_I^{ij}\,,\,  \\
& & \qquad m,n,p=0,\ldots,5\,;\quad a,b=1,..., 4\,;\quad
i,j=1,2\,;\quad I=1,2,3\nonumber\end{eqnarray}
where
$\epsilon^{12}=- \epsilon_{12}=1$, and $\tau^{ij}_I$ ($I=1,2,3$)
provide a basis of $2\times 2$ symmetric matrices and are
expressed through the usual $SU(2)$ group Pauli matrices, $\tau_{I
\, ij}= \epsilon_{jj^\prime}\, \sigma_{I\, i}{}^{j^\prime}$.
  The matrices
${\tilde \gamma}^{ab}_m$ ($\gamma_{ab}^m= 1/2 \,
\varepsilon_{abcd} {\tilde \gamma}^{m\, cd}$) provide a complete
set of $4\times 4$ antisymmetric matrices with upper (lower)
indices transforming under an (anti)chiral fundamental
representation of the non--compact group $SU^*(4)\sim Spin(1,5)$.
For the space of $4\times 4$ symmetric matrices with upper (lower)
indices
  a basis is provided by  the set of self--dual
(anti--self--dual) matrices   $({\tilde \gamma}^{m n p})^{ab}$
[$\gamma_{ab}^{m n p}$],
\begin{equation}\label{ggsd}
({\tilde \gamma}^{m n p})^{ab}={1\over 3!}\epsilon^{mnpqrs}{\tilde
\gamma}^{ab}_{qrs}\; ,  \qquad \gamma_{ab}^{m n p}=- {1\over
3!}\epsilon^{mnpqrs}(\gamma_{qrs})_{ab}\; .
\end{equation}
The coordinates
$ x^m=\,x^{ai\,bj}\,\gamma^m_{ab}\,\epsilon_{ij}$
 are associated with $D=6$ space--time, while the
self-dual coordinates \begin{equation}\label{6Dy=asd} y_I^{mnp}=
x^{ai\,bj}\,\gamma^{mnp}_{ab}\,\tau_{I\,ij}=- {1 \over 3!
}\,\epsilon^{mnpqrs}y^I_{qrs}\; , \end{equation} describe spinning
degrees of freedom.

The discussion repeats again that of the case for  $D=4$ and $D=10$.
Because of the form of the tensorial coordinates in (\ref{6})
 the six dimensional analogue
of the expansions (\ref{is}) and (\ref{his}) contains powers
of $y_i^{mnp}$. Corresponding field strengths, which again describe conformal fields in six dimensions, are
 self--dual with respect to each set of the indexes $(m_i n_i p_i)$.
 In other words, one has an infinite number of conformally
invariant (self-dual) ``multi--3--form" higher spin fields in the six--dimensional
space--time  which form
the $(2[s]+1)$-dimensional representation of the group $SO(3)$.

It is possible to describe  interactions with higher spin currents in terms of tensorial spaces.
In order to do so (we shall describe a main idea and refer to \cite{Gelfond:2010pm}, \cite{Vasiliev:2007yc}--\cite{Gelfond:2003vh}
for the detailed discussion and derivations)
 the equation (\ref{Y}) has been generalized  in \cite{Gelfond:2003vh} to include several commuting variables
\begin{equation}\label{Yrr}
\left({\partial\over{\partial
X^{\alpha \beta}}} \pm i\eta^{ij}{\partial^2\over{\partial Y^{i \alpha} \partial
Y^{j \beta}}}\right)C^r_\pm(X,Y)=0.
\end{equation}
where $i,j=1,...,r$ and $\eta^{ij}=\eta^{ij}$ is a nondegenerate metric.
As we explained above  free higher spin fields in $D=4$ are described by the rank one
equations in the ten dimensional tensorial space.
On the other hand, higher spin currents are fields of rank two $r=2$.
These currents obey the equations with off--diagonal $\eta^{ij}$ \cite{Gelfond:2008ur}.
These currents $J(X,Y^i)$ are bilinear in the higher spin gauge fields
${\cal C}_+$ and   ${\cal C}_-$,  which obey rank one equation
(\ref{Yrr})
$J={\cal C}_+{\cal C}_- $.

On the other side when considering rank two equations the corresponding tensorial
space can be embedded in the higher dimensional tensorial space.
From the discussion above it follows that a natural candidate for such higher dimensional space is tensorial extension
of $D=6$ space--time.
In this way one effectively linearizes the problem since conformal currents in four dimensions are identified
with the fields in $D=6$ \cite{Gelfond:2010pm}.

\subsection{AdS space-time}

A discussion for tensorial extension of $AdS_4$ spaces follows the same lines as for tensorial extension
of Minkowski space--time \cite{Didenko:2003aa}--\cite{Plyushchay:2003tj}. In order to better explain
how a tensorial extension of anti de Sitter space--time looks let us return for a moment  to the four dimensional
Minkowski space--time
and consider the symmetries of the equation (\ref{PRE}).
We shall concentrate mainly on four space--time dimensions.
In accordance with the conjecture of \cite{Fronsdal:1985pd} the equations (\ref{PRE}) are invariant under
the transformations of the $Sp(8)$ group \cite{Plyushchay:2003gv}
\begin{equation}\label{trr}
\delta\lambda_\alpha=g_\alpha^{~\beta}\lambda_\beta-k_{\alpha \beta}X^{\beta \gamma}, \quad
\delta X^{\alpha \beta} =
a^{\alpha \beta}-
(X^{\gamma \beta}g_\gamma^{~\alpha}+X^{\gamma \alpha}g_\gamma^{~\beta})+X^{\alpha \gamma}k_{\gamma \delta}X^{\delta \beta}.
\end{equation}
The parameters $a^{\alpha \beta}$, $g_\gamma^{~\alpha}$ and  $k_{\alpha \beta}$ correspond to the generators
$P_{\alpha, \beta}$ (translations) $G_\beta^{~\alpha}$ ($GL(4)$ group) and $K_{\alpha \beta}$ (conformal boosts)
 of the $Sp(8)$ group\footnote{Let us note that $Sp(8)$ algebra can be conveniently realized after introducing
a twistor variable, conjugate to $\lambda_\alpha$ as $[ \mu^\alpha, \lambda_\beta,]= \delta^\alpha_\beta$. The generators of the $Sp(8)$
group will then have a form $P_{\alpha \beta}=\lambda_\alpha \lambda_\beta$, $G_\alpha^{~\beta}= \lambda_\alpha \mu^\beta$,
$K_{\alpha \beta}=\mu_\alpha \mu_\beta$.}
$$
[P,P]=0\,,\quad [K,K]=0,
$$
\begin{equation}\label{15b}
 [P,K]\sim G\,, \quad [G,G]\sim
G\,,
\quad [G,P]\sim P\,, \quad [G,K]\sim K\,.
\end{equation}
Let us note that under the transformations (\ref{trr}) one has
\begin{equation}\label{20.1}
\delta dX^{\alpha \beta}= dX^{\alpha'\beta}\,g_{\alpha'}^{~~\alpha}(X)\,+\,
 dX^{\alpha\beta'}\,g_{\beta'}^{~~\beta}(X)\,, \quad
\end{equation}
where $g_{\alpha'}^{~\alpha}(X)$ are the infinitesimal
$Sp(8)$ transformations nonlinearly realized on the coset
superspace ${{Sp(8)}\over{GL(4)\times\!\!\!\!\supset K}}$ in
terms of $GL(4)$ matrices.
 We have thus demonstrated that the
tensorial extension of flat four dimensional Minkowski space  is a coset
space ${{Sp(8)}\over{GL(4)\times\!\!\!\!\supset K}}$, and $e^{X_{\alpha \beta} P^{\alpha \beta}}$
is a coset element.

A tensorial extension of $AdS_4$ space is just another coset of the $Sp(8)$ group.
Recall that the usual $AdS_4$ space is a coset space
$SO(2,4)\over{(SO(1,3)\times D) \times\!\!\!\!\supset K}$
parametrized by the  coset element $e^{{\cal P}_m\,x^m}$.
 The generators of the
$AdS_4$ boosts can be singled out from the generators of the four
dimensional conformal group $SO(2,4)$ by taking a linear
combination of the generators of Poincar\'e translations $P_m$ and
conformal boosts $K_m$ as ${\cal P}_m=P_m+K_m$.

Analogously, for the case of tensorial extension of $AdS_4$ space
let us consider the generators
\begin{equation}\label{6.1}
 {\cal P}_{\alpha \beta}= P_{\alpha \beta}+K_{\alpha \beta}, \quad [{\cal P},{\cal P}]\sim
M,
 \quad [{\cal P},M]\sim {\cal P},
\end{equation}
 where $M_{\alpha \beta}$ stands for $G_{\{\alpha \beta \}}=G_{\{\beta \alpha \}} $.
 One can see that the corresponding manifold is an $Sp(4)$ group manifold
 \cite{Plyushchay:2003gv} which can be realized
as a coset space ${{Sp(8)}\over{GL(4) \times\!\!\!\!\supset K}}$ with the coset
element $e^{(P+K)_{\alpha \beta}\,X^{\alpha \beta}}$.

The consideration of the corresponding free field equations on tensorial  space is simplified
by the fact, that vielbiens (Cartan forms)
$\omega^{\alpha
\beta}=dX^{\mu\nu}\,\omega^{~~\alpha\,
\beta}_{\mu\nu}(X)$
on the $Sp(2n)$ group manifold differ
from the ones of the flat manifold by $GL(2n)$ rotations \cite{Plyushchay:2003gv}
\begin{equation}  \label{ANB}
\omega^{\alpha \beta}(X) = d X^{\alpha^\prime
\beta^\prime}\;G_{\alpha^\prime}{}^\alpha\,(X)
G_{\beta^\prime}{}^\beta\,(X), \quad G_\beta^{-1\alpha}(x)=\delta^{~\alpha}_\beta+{\varsigma\over 4}x^{~\alpha}_\beta,
\end{equation}
where the parameter $\varsigma$ can be associated with the inverse of $AdS_4$ radius in case of $Sp(4)$ group.
This property of ``$GL(2n)$ flatness" of $Sp(2n)$ group manifolds, which is a generalization of conformal flatness of usual
AdS spaces, can be established by
checking that the expression (\ref{ANB}) solves
 Maurer--Cartan equations for the $Sp(2n)$
group
\begin{equation}\label{mce}
d\omega^{\alpha \beta}+{\varsigma\over 2}\omega^{\alpha \gamma}\wedge
\omega_\gamma^{~\beta}=0.
\end{equation}
Further, one has to consider an $Sp(4)$ counterpart of the equation (\ref{PRR}).
Recall that for the case of the tensorial extension of Minkowski space
one has
 $P_{\alpha \beta}= i \frac{\partial}{ \partial X^{\alpha \beta}}$.
Similarly, for the tensorial extension of $AdS_4$ one has
${\cal P}_{\alpha \beta}= i \nabla_{\alpha \beta}$, where due to the
  the $GL(4)$ flatness of the $Sp(4)$ group manifold one has
$\nabla_{\alpha \beta}=G_\alpha^{-1\gamma}(x)G_\beta^{-1\delta}(x)\frac{\partial}{\partial X^{\gamma \delta}}$.
 The operators $\nabla_{\alpha \beta}$ form $Sp(4)$ algebra.
Therefore the $Sp(4)$ version of the equation (\ref{PRR})
will have the form
\begin{equation}\label{lsp}
\left[\nabla_{\alpha \beta}-{i\over 2}(Y_\alpha
Y_\beta +Y_\beta Y_\alpha)\right]\Phi(X,\lambda)=0\,, \quad Y_\alpha\equiv
\lambda_\alpha+{i\varsigma\over 8} {\partial\over {\partial\lambda^\alpha}}\,.
\end{equation}
The reason for the appearance of the terms which are proportional to $\varsigma$ in the equation (\ref{lsp})
is that these terms ensure that the second term in (\ref{lsp})  obeys $Sp(4)$ algebra, as the derivatives
$\nabla_{\alpha \beta}$ do.
An $Sp(4)$ version of the equation (\ref{Y}) now looks like
\begin{equation}\label{ysp}
\left[\nabla_{\alpha \beta}-{i\over 2}(Y_\alpha
Y_\beta+Y_\beta Y_\alpha)\right]C(X,Y)=0,\,\quad Y_\alpha\equiv {i}
{\partial\over {\partial y^\alpha}}+{\varsigma\over 8}y_\alpha\,.
\end{equation}
The equations (\ref{lsp})--(\ref{ysp}) can be solved to obtain ``plane wave" solutions
for  tensorial $AdS_4$ spaces
\begin{equation}\label{ls}
\Phi(X,\lambda)  =\int\, d^4y\, \sqrt{\det G^{-1}(x)}\,
e^{{i}X^{\alpha \beta}(\lambda_\alpha+ {\varsigma\over 8}y_\alpha
)(\lambda_\beta+{\varsigma\over 8}y_\beta) +i \lambda_\alpha y^\alpha}\,\varphi(y)\,,
\end{equation}
\begin{equation}\label{ys}
C(X^{\alpha \beta},y)=\int\, d^4\lambda \, \sqrt{\det G^{-1}(x)}\,
e^{{i}x^{\alpha \beta}(\lambda_\alpha+ {\varsigma\over 8}y_\alpha
)(\l_\beta+{\varsigma\over 8}y_\beta) -i \lambda_\alpha y^\alpha}\,\varphi(\lambda)\,.
\end{equation}
Finally, after some algebra one can find equations which are analogous to (\ref{b})
\begin{eqnarray}\label{bsp}
\nabla_{\alpha [\beta}\nabla_{\gamma]\delta}b(x)&=&{\varsigma\over
16}\left(C_{\alpha [\beta }\nabla_{\gamma]\delta}- C_{\delta [\gamma}\nabla_{\beta] \alpha} +
2C_{\beta \gamma}\nabla_{\alpha \delta}\right)b(x)+ \\ \nonumber
&&{\varsigma^2\over
64}\left(2C_{\alpha \delta}C_{\beta \gamma}-C_{\alpha [\beta}C_{\gamma]\delta}\right)b(x),
\end{eqnarray}
\begin{equation}\label{fsp}
\nabla_{\alpha[\beta }f_{\gamma]}(x)=-{\varsigma\over
4}\left(C_{\alpha [\gamma}f_{\beta]}(x)+2C_{\beta \gamma}f_\alpha(x)\right)\,,
\end{equation}
where $C_{\alpha \beta}$ is a four--dimensional charge conjugation matrix.

In order to obtain free unfolded equations on $AdS_4$ one
multiplies the equation (\ref{ysp}) with
${1\over 2}
G^{~\alpha}_{\delta}\,\gamma_m^{\delta \sigma}G_{\sigma}^{~\beta}$
and then takes $y^{mn}=0$ to obtain
\begin{equation}\label{adssp}
\left[{\partial\over{\partial
x^m}}+i(
{1\over 4}dx^m\,\omega_m^{ab}(x)\,\gamma_{ab}^{\alpha\beta}+{1\over 2}
dx^m\,e_m^a(x)\gamma_a^{\alpha \beta})
Y_\alpha \,Y_\beta\right]C(x^m,y^\alpha)=0.
\end{equation}
Let us note that instead of   (\ref{ys}),
one could have used another form of the generating functional
 \cite{Bolotin:1999fa}, \cite{Plyushchay:2003gv},
which has been obtained for the
 conformally flat parametrization of
$AdS_4$. Again, repeating similar steps as  in the Subsection 2.1
for the case of flat Minkowski space--time, one can obtain free unfolded equations for massless higher spin fields
on $AdS_4$ \cite{Vasiliev:1990en}.

\setcounter{equation}0\section{Cubic and Higher Order Interactions}
\subsection{Cubic vertices on a flat background}\label{Cubicf}
In this section we briefly recall the BRST method for the construction of  cubic interaction vertices
on a flat background (a more detailed review as well as generalization to AdS backgrounds
can be found in \cite{Fotopoulos:2008ka}, \cite{Buchbinder:2006eq}).
This method is quite general since it can be used for the construction of nonabelian
\cite{Bengtsson:1987jt},  \cite{Fotopoulos:2010ay}, \cite{Fotopoulos:2007yq}
 and abelian \cite{Koh:1986vg}, \cite{Fotopoulos:2007nm}  interaction vertices.
 Essentially the method of BRST constructions is nothing else but
 a method for constructing gauge invariant Lagrangians, where extra
 gauge degrees of freedom are introduced in order to ensure gauge invariance
for unconstrained gauge fields and gauge transformation parameters.

Let us recall that free massless fields with a spin $s$ on a $D$ -- dimensional flat background satisfy mass-shell and transversality
conditions.
 \begin{equation}\label{condi}
  \partial^\mu \partial_\mu \phi_{\mu_1,.., \mu_s}(x)=  \partial^{\mu_1}  \phi_{\mu_1,.., \mu_s}(x)=0.
  \end{equation}
In case one describes an irreducible higher spin mode one has to add to (\ref{condi})  an extra equation $\phi^{\mu_1}{}_{\mu_1,.., \mu_s}(x)=0$.
Without this zero trace condition the equations (\ref{condi}) describe fields with spins $s,s-2,...,1/0$ simultaneously.

In order to construct a free Lagrangian which gives (\ref{condi}) as a result of equations of motion
one can introduce an auxiliary Fock space spanned by oscillators
 $[\alpha_\mu , \alpha_\nu^+]=\eta_{\mu \nu}$, with the vacuum defined as $\alpha_\mu |0 \rangle_\alpha =0 $.
Further, let us introduce operators $l_0=-\partial^\mu \partial_\mu$ (D`Alembertian), $l = -i\alpha^\mu \partial_\mu$ (divergence)
 as well as its hermitian conjugate
 $l^+ =-i \alpha^{\mu+} \partial_\mu$.
  The BRST charge   for this set of operators is very simple since the only nonzero commutator is
 $[l, l^+]= l_0$.
  Introducing anticommuting ghost variables $c_0 ,c, c^+$
  (with ghost number $+1$), conjugate momenta $b_0, b, b^+$ (with ghost number $-1$), with anticommutation relations
  $ \{c_0, b_0 \} = \{c, b^+ \}= \{c^+, b \}=1 $ as well as ghost vacuum $b_0|0 \rangle_{gh}=b|0 \rangle_{gh}=c|0 \rangle_{gh}=0$
  and $|0 \rangle= |0 \rangle_{\alpha} \otimes |0 \rangle_{gh}$
  one can construct a nilpotent BRST charge\footnote{It is possible to make an AdS deformation of the flat BRST charge
  \cite{Buchbinder:2001bs}--\cite{Sagnotti:2003qa}. The cohomologies of these BRST charges reproduce free field equations on AdS \cite{Metsaev:1997nj}.}
  \begin{equation}\label{BRST}
  Q= c_0 l_0 + c l^+ + c^+ l - c^+ c b_0,  \quad Q^2=0.
  \end{equation}
Solving cohomologies of the BRST charge (\ref{BRST}) one can obtain conditions (\ref{condi}) on the physical field.
Because of the presence of ghost variables one has extra degrees of freedom in the system, however
the gauge fixing procedure
entirely removes these extra degrees of
freedom and leads to the equations (\ref{condi}) with residual gauge invariance
$\delta |\phi \rangle = l^+ |\lambda \rangle $, $l_0|\lambda \rangle=l|\lambda \rangle=0$

In order to consider cubic interactions one needs to take three copies of the Hilbert space described above.
The BRST charge is a sum of ``individual'' ones
$Q=Q_1+ Q_2 + Q_3$ and  is  nilpotent since  $\{Q_i, Q_j \} =0$.
 A cubic Lagrangian
 \begin{equation} \label{cubic}
 L= \sum_{i=1,2,3} \int dc_0^i \langle \Phi_i| Q_i| \Phi_i  \rangle +
 g (\int dc_0^1 dc_0^2 dc_0^3 \langle \Phi_1| \langle \Phi_2| \langle \Phi_3||V \rangle + h.c. )
 + O(g^2)
 \end{equation}
 is invariant up to terms of order $g$ under the transformations
 \begin{equation}\label{trrr}
 \delta | \Phi_i  \rangle = Q_i | \Lambda_i  \rangle - g \int  dc_0^{i+1} dc_0^{i+2} ((\langle \Phi_{i+1}| \langle \Lambda_{i+2}|+
\langle \Phi_{i+2}| \langle \Lambda_{i+1}| )    |V \rangle) + O(g^2)
\end{equation}
provided the vertex operator  $| V \rangle$ is BRST invariant
\begin{equation}\label{BRSTINV}
Q |V \rangle =0.
\end{equation}
The procedure of solving of the equation (\ref{BRSTINV}) is simplified after taking
two points  into account.
From the Lagrangian (\ref{cubic}) one can conclude that the vertex $|V \rangle$ has the ghost number three,
so writing it in the form
\begin{equation}
|V \rangle=V  c_0^1c_0^1c_0^1 |0\rangle_1 \otimes  |0\rangle_2 \otimes |0\rangle_3 \equiv V|0\rangle_{123}
\end{equation}
one can conclude that the function $V$ has the ghost number zero. There are only five Lorentz invariant combinations
with ghost number zero: $p^{\mu i} p_\mu^j$, $\alpha^{\mu i+} p_\mu^j$, $\alpha^{\mu i+} \alpha^{j+}_\mu$,
$c^{i+}b^{j+}$ and $c^{i+}b_0^j$; where $p_\mu= -i \partial_\mu$ and the index $i=1,2,3$ numerates separate Hilbert spaces.
Therefore, in general the unknown function $V$ is a series expansion in terms of these
combinations with unknown coefficients and the BRST invariance condition (\ref{BRSTINV}) fixes these coefficients.

The second point is that the number operator $N$, which counts the total number of oscillators
$\alpha^{\mu i+}$, $c^{i+}$ and  $b^{i+}$
\begin{equation} \label{OPER}
N= \alpha^{\mu i+} \alpha^{i}_\mu + c^{i+}b^{i}+ b^{i+}c^{i},
\end{equation}
commutes with the BRST charge $Q$.
Therefore, the equation (\ref{BRSTINV}) splits into separate equations according to eigenvalues of the operator (\ref{OPER}).
Finally, one needs to discard BRST trivial vertices, because they correspond to the vertices which
can be obtained from the free Lagrangian via field redefinitions. One
 also has to take into account the vanishing of the total derivative $p_\mu^1 + p_\mu^2+ p_\mu^3=0$.

After these preliminary remarks we can consider cubic interaction vertices for massless higher spin fields
on the flat background  in the BRST approach.
Let us first consider  the solution given in
 \cite{Bengtsson:1987jt} in detail.
On the first level i.e., for the vertex with total number of oscillators $N=1$ one has the expansion
\begin{equation}\label{1}
\Delta_1 = Y_{ij} \alpha^{i+}_\mu p_\mu^j + Z_{ij}
c^{i+}b_0^j,
\end{equation}
on the second level $N=2$
\begin{equation} \label{2}
\Delta_2= S_{ij} c^{i+} b^{j+} + \frac{P_{ij}}{2} \alpha^{i+}_\mu \alpha^{j+}_\mu
\end{equation}
and for  $N=3$
\bea
\label{Bansatz2} \nonumber
 \Delta_3& =& {\tilde X}^{(1)}_{rstu} (\alpha^{r+}_\mu
\alpha^{s+}_\mu) (\alpha^{t+}_\nu p_{\nu}^u)  +
{\tilde X}^{(2)}_{rstu} (c^{r+} b^{s+}) (\alpha^{t+}_\mu p_{\mu}^u) + \\
&&{\tilde X}^{(3)}_{rstu} (\alpha^{r+}_\mu \alpha^{s+}_\mu) (c^{t+} b_0^u)
+ {\tilde X}^{(4)}_{rstu} (c^{r+} b^{s+}) (c^{t+} b_0^u). \eea
The coefficients ${ \tilde X}^{(1)}_{rstu}, {\tilde X}^{(2)}_{rstu}, {\tilde X}^{(3)}_{rstu}$
and ${\tilde X}^{(4)}_{rstu}$ obey the symmetry relations \be \label{SIMS}
{\tilde X}^{(1)}_{rstu}= {\tilde X}^{(1)}_{srtu}, \quad {\tilde X}^{(3)}_{rstu}=
{\tilde X}^{(3)}_{srtu}, \quad  {\tilde X}^{(5)}_{rstu}= - {\tilde X}^{(5)}_{tsru} \ .
\ee
Let us note \cite{Buchbinder:2006eq} that the expressions for $\Delta_1$ , $\Delta_2$ and $\Delta_3$
do not contain powers of $p^{\mu i} p_{\mu}^j$ since this kind of terms would belong to trivial cohomologies
of the BRST charge (\ref{BRST}).
The BRST invariance conditions  for  $\Delta_1$,
$\Delta_2$ and $\Delta_3$  are independent from each other since they have different values of $N$.
 For (\ref{1}) and (\ref{2})
one gets
\begin{equation}\label{KOsolution}
Z_{i,i+1}+Z_{i,i+2}=0
\end{equation}
$$
Y_{i,i+1}= Y_{ii}-Z_{ii} -1/2(Z_{i,i+1}-Z_{i,i+2})
$$
$$
Y_{i,i+2}= Y_{ii}-Z_{ii} +1/2(Z_{i,i+1}-Z_{i,i+2}).
$$
\begin{eqnarray}\label{KOmsol}
&& S_{ij}= P_{ij}=0, \qquad i\neq j, \\
&& P_{ii} - S_{ii}=0, \qquad i=1,2,3 \nonumber
\end{eqnarray}
Notice that one can take diagonal elements of $Y_{ij}$ and $Z_{ij}$ equal to zero. Imposing a cyclic symmetry
one can see that the solution (\ref{KOsolution}) is expressed via a single parameter.
Similarly, after imposing  cyclic symmetry one can see that the solution (\ref{KOmsol})
is expressed via another single  parameter.
The BRST invariance condition for (\ref{Bansatz2}) implies
\be \label{nontr1}
(2 {\tilde X}^{(1)}_{rstu} p_\mu^r p_\nu^u -
{\tilde X}^{(2)}_{rstu} p_\mu^s p_\nu^u ) c^{r+} \alpha^{s+}_\mu
\alpha^{t+}_\nu=0,
\ee
\be \label{nontr2}
 (- {\tilde X}^{(3)}_{rstu} p_\mu^u p_\mu^u +
{\tilde X}^{(1)}_{rstu} p_\mu^t p_\mu^u ) c^{t+} \alpha^{r+}_\nu
\alpha^{s+}_\nu=0,
\ee
\be \label{nontr3}
(- {\tilde X}^{(5)}_{rstu} p_\mu^u p_\mu^u +
{\tilde X}^{(2)}_{rstu} p_\mu^t p_\mu^u ) c^{r+} b^{s+}c^{t+}=0,
\ee
\be \label{nontr4}
(-
{\tilde X}^{(2)}_{rtsu} b_0^t p_\mu^u +2 {\tilde X}^{(3)}_{rstu} b_0^u   p_\mu^r -
{\tilde X}^{(5)}_{rstu} b_0^u   p_\mu^s ) c^{r+} c^{t+} \alpha^{s+}_\mu =0,
\ee
\be \label{nontr5}
 {\tilde X}^{(5)}_{rstu}b_0^s c^{s+} c^{r+} c^{t+}b_0^u=0.
\ee
After imposing  cyclic symmetry on the coefficients
${\tilde X}^{(1)}_{rstu}$ under the indexes $r,s,t,u$ these equations  can
be solved  \cite{Bengtsson:1987jt} to give the solution given in
Table 1 where all entries are proportional to yet another free paramater.

\begin{table}
\begin{center}
\begin{tabular}[t]{|c|c|c|c|c|}%\trule
Index combination & ${\tilde X}^{(1)}$ & ${\tilde X}^{(2)}$ & ${\tilde X}^{(3)}$ &
${\tilde X}^{(4)}$ \\ %\trule
1231 & 1 & -2 & -1 &  1 \\ %\trule
1232 & -1 & 0 & 1 & -1 \\ %\trule
1233 & 0 & 0 & 0 &  1 \\ %\trule
1211 & 0 & 0 & 1 &  0 \\ %\trule
1212 & -1 & 2 & 0 &  0 \\ %\trule
1213 & -1 & 2 & 0 &  0 \\ %\trule
1221 & 1 &-2 & 0 &  -4 \\ %\trule
1222 & 0 &0 & -1 &  1 \\ %\trule
1223 & 1 &0 & 0 & -1 \\ %\trule
1111 & 0 &0 & 0 & 0 \\ %\trule
1112 & -1 &-2 & 1 & 0 \\ %\trule
1113 & 1 &2 & -1 & 0 \\ %\trule
1121 & -2 &-4 & 1 & 1 \\ %\trule
1122 & -6 &-12 & -5 & -4 \\ %\trule
1123 & 0 &-2 & -1 & -1 \\ %\trule
1131 & 2 &4 & -1 & -1 \\ %\trule
1132 & 0 &2 & 1 & 1 \\ %\trule
1133 & 6 &12 & 5 & 4 \\ %\trule
2131 &  & 0 &  &  \\ %\trule
2132 &  & 2 &  &  \\ %\trule
2133 &  & 0 &  &  \\ %\trule
2111 &  & 0 &  &  \\ %\trule
2112 &  & 2 &  &  \\ %\trule
2113 &  & 0 &  &  \\ %\trule
2121 &  & -2 &  &  \\ %\trule
2122 &  & 0 &  &  \\ %\trule
2123 &  & -2 &  &  \\ %\trule
\end{tabular}
\end{center}
\caption{Empty entries in the table mean that the corresponding
value of the coefficient can be recovered from the ones given in
the table using the cyclic property of indices (for example
${\tilde X}^{(i)}_{1231}= {\tilde X}^{(i)}_{2312}={\tilde X}^{(i)}_{3123}$) and symmetry
properties (\ref{SIMS})} \label{tabSpectra}
\end{table}
As  shown in \cite{Bengtsson:1987jt} the solution given in  Table 1 reproduces a usual Yang--Mills vertex
for three spin one fields provided, the fields are equipped with gauge (Chan --Paton) indexes.
However, the solution obtained in \cite{Bengtsson:1987jt} has many more implications. In particular, taking an exponential of this solution
i.e., considering
\begin{equation} \label{TOTAL}
{\cal V} |0\rangle_{123} = e^{\Delta_1 + \Delta_2 + \Delta_3} |0\rangle_{123}
\end{equation}
results in an of--shell extension of the vertex given in \cite{Sagnotti:2010at}
(see also \cite{Fotopoulos:2010ay}, \cite{Taronna:2011kt} for relevant discussion).
Again the dependence of the vertex (\ref{TOTAL}) on  $p^{\mu i}p_{\mu}^j$
can be removed via field redefinitions and the BRST invariance conditions \cite{Buchbinder:2006eq}.

The solution containing only a part with $\Delta_1$ and $\Delta_2$ is discussed in detail
in \cite{Fotopoulos:2007nm}. It has been shown that choosing $S_{ii}=1$ makes the theory purely cubic,
 because four point functions are zero in this case. Therefore the vertex which contains
only $\Delta_1$ and $\Delta_2$ gives trivial interactions beyond the cubic level.

Let us note that cubic vertices on flat space--time might have yet another interesting application.
In particular, they can be deformed to obtain interaction vertices on $AdS$
background  \cite{Vasilev:2011xf}. A construction of vertices on an AdS
background is a very interesting subject in its on right
 \cite{Buchbinder:2006eq}-- \cite{Fotopoulos:2007yq}, \cite{Vasilev:2011xf}--\cite{Joung:2011ww}
 and it can also be  relevant for
studies of AdS/CFT correspondence (see for example \cite{Koch:2010cy}--\cite{Campoleoni:2010zq})
which is, however, beyond the scope of the present review.

\subsection{A short review of the BCFW method}\label{BCFWrev}

The key point of BCFW method \cite{Britto:2004ap}--\cite{Britto:2005fq}, which is based upon twistor formulation
of gauge theory \cite{Witten:2003nn}, is that tree level
amplitudes constructed using Feynman rules are rational functions
of external momenta. Analytic continuation of these momenta on the
complex domain  turns the amplitudes into meromorphic functions
which can be constructed solely by their residues. Since the
residues of scattering amplitudes are, due to unitarity, products
of lower point on-shell amplitudes the final outcome is a set of
powerful recursive relations.

 The simplest complex deformation involves only two external  momenta
 \begin{equation}\label{shift}
 \hat p_i(z)= p_i - q z\ , \quad
 \hat p_j(z)= p_j + q z \ ,
\end{equation}
where $z$ is a complex variable and $q \cdot p_{i} =q\cdot p_j= 0$,   $q^2 = 0$.
In Minkowski
space-time this is only possible for complex $q$.
 As  discussed in the Section 2 in four dimensions one can use spinor representations of momenta
 (\ref{BW}).
BCFW shift (\ref{shift}) corresponds to the following shift in the commuting spinors
\begin{equation}
\hat{\lambda}_A^{(i)}(z) = \lambda_A^{(i)} + z\lambda_A^{(j)}, \quad
\hat{\tilde\lambda}_{\dot{A}}^{(j)}(z) = \tilde\lambda_{\dot{A}}^{(j)} -
z\tilde\lambda_{\dot{A}}^{(i)}.
\end{equation}
A general amplitude after the BCFW shift becomes a mereomorphic function of a complex
variable $z$. Simple poles of the function correspond to the values of $z$ where the propagators of intermediate states go
on--shell on the complex plane.
Finally, an undeformed amplitude can be computed using Cauchy's theorem
\begin{equation}
{\cal M}_n (0) = {1\over 2\pi i}\oint_{z=0} \frac{{\cal M}_n(z)}{z} dz = -
\left\{\sum \mathrm{Res}_{z=\textrm{finite}} + \mathrm{Res}_{z=
\infty} \right\}. \
\end{equation}

In four dimensions polarization vectors can be also represented in terms of commuting spinors
\begin{eqnarray}\label{spinor}
&&\epsilon^+_{A\dot A} = \frac{\mu_A\tilde\lambda_{\dot
A}}{\langle \mu ,\lambda \rangle},  \qquad  \epsilon^-_{A\dot A}
=\frac{\lambda_A\tilde\mu_{\dot A}}{[\tilde\lambda , \tilde\mu]} \\
&& \langle \mu ,\lambda \rangle \equiv \mu_A \lambda_B
\epsilon^{AB} \qquad [\tilde\lambda , \tilde\mu] \equiv
\tilde\mu_{\dot A} \tilde\lambda_{\dot B} \epsilon^{\dot A \dot
B}\nonumber
\end{eqnarray}
with $\mu_A$ and $\tilde\mu_{\dot A}$ arbitrary reference spinors.
Polarizations of higher spin states  are given by products of the
polarizations for spin one
\begin{equation}\label{polten}
\epsilon^+_{A_1 \dot{A}_1 \dots A_s \dot{A}_s} = \prod_{i=1}^s
\epsilon ^+_{A_i \dot{A}_i} \qquad \epsilon^-_{A_1 \dot{A}_1 \dots
A_s \dot{A}_s} = \prod_{i=1}^s \epsilon ^-_{A_i \dot{A}_i}.
\end{equation}

In order for the BCFW recursion relation to be applicable the deformed
amplitude should vanish when  $z \to \infty$. This is because a pole at infinity,
unlike the ones for finite values of $z$, does not have an interpretation
as product of lower point amplitudes.
In \cite{Benincasa:2007xk} a very interesting criterion was
derived in order to classify, in four dimensions, which theories
are constructible (have zero pole at infinity) under BCFW deformations. The criterion has been
stated explicitly for the four-point function and it is a
necessary condition for a theory to have zero residue at infinity.
  Say we denote by ${\cal
M}^{(i,j)}(z)$ the four-point function under deformation of
particles $i$ and $j$. Assume further  that the helicities $h_2$
and $h_4$ are negative while $h_1$ is positive.  The criterion
advocates that
\begin{equation}\label{BC1}
{\cal M}_4^{(1,2)}(0)={\cal M}_4^{(1,4)}(0).
\end{equation}
Note that in $M^{(1,2)}(z)$ only poles
from the $t,\ u$ channels on the complex z-plane will contribute
since the expression $1/(p_1(z)+p_2(z))^2=1/ (p_1+p_2)^2$ is
$z$--independent and it does not give a pole. Similarly
 for $M^{(1,4)}(0)$ only poles from the $s, \ u $
channels on the complex-z plane will contribute.\footnote{We use the following definition for Mandelstam variables
$s={(p_1+p_2)}^2$, $t={(p_1+p_4)}^2$,
$u={(p_1+p_3)}^2$.}  So the
crossing symmetry condition (\ref{BC1}) is a highly nontrivial
constraint.

\subsection{BCFW relations and higher spin fields}

BCFW recursion relations have wide applications in Yang--Mills gauge theories and
supergavities (see \cite{Feng:2011gc} for a  review), whereas
the application of BCFW relations to string theory
have been discussed relatively recently \cite{Cheung:2010vn} -- \cite{Boels:2010bv},
 \cite{Fotopoulos:2010cm}--\cite{Fotopoulos:2010jz} (see also \cite{Schlotterer:2010kk}--\cite{Boels:2012if}
  for interactions between  higher spin fields in string theory).
The main ingredient in this consideration is the Pomeron vertex operator introduced in
\cite{Brower:2006ea}. This vertex operator appears when considering the
Operator Product Expansion(OPE) between string vertex operators in a Regge regime
$s \gg t$  (a high energy limit). The
Pomeron vertex operator can be obtained from OPE for usual string vertex operators in a Regge regime
$s \gg t$
 \begin{equation}
{\cal V}_P \sim  C_n\ \Gamma\left(\alpha^\prime p^2 -1\right) \ e^{ip\cdot X}[ q\cdot
\partial X]^{1- \alpha^\prime p^2 }
\end{equation}
where $p=p_1+p_2$ and $q=p_1-p_2$.
The operator $C_n$ depends on the states for which we consider OPE and is a function of their
polarizations and momenta.
 Pomerons are physical composite states, but with a fractional oscillator number, and therefore they are outside a normal Hilbert space
 \begin{equation}
 L_0 {\cal V}_{P}=0, \quad  L_1 {\cal V}_{P}=0.
\end{equation}
 Taking into account Pomeron vertex operators when considering four point functions, say for a bosonic string, one can show
that the typical behavior under BCFW shift is
\begin{equation}
{\cal M}(z) \sim z^{n+1+ \alpha^\prime P_{12}^2}
\end{equation}
i.e., open bosonic string amplitudes are constructible in the regime
$n+1+\alpha^\prime  P_{12}^2<0$.
The inclusion of the Pomeron vertex operator when considering BCFW shift is natural.
Indeed the Pomeron vertex operator appears
at the Regge (high energy) limit of the bosonic string and, on the other hand, BCFW shift of the momenta
gives as high energy limit as we take $z \to \infty$.

Now one can try to apply similar arguments for massless higher spin theories.
Below we give only some brief arguments and refer to \cite{Fotopoulos:2010ay} for  a detailed discussion.
Let us recall that free field equations can be recovered from
the usual BRST charge for the bosonic string by taking $\alpha^\prime \to \infty$
limit (see, for example, \cite{Sagnotti:2003qa}) and  one can further consider interactions in this limit \cite{Sagnotti:2010at}.
Therefore, when considering BCFW relations for these theories it is natural to also include  Pomeron vertices
in the theory (which are relevant to the tensile bosonic string theory in the high energy limit) and which make the theory constructible.
Let us also note  that in the high energy limit the Pomeron vertex satisfies the physical state conditions.

 As a spin of intermediate fields  grows a dimension of the coupling constant becomes smaller and relevant diagrams become more divergent.
Therefore, the simplest system which includes a coupling of a massless higher spin field with two scalar fields
contains a lot of information i.e, it allows us to indicate a problem.
Below we give two explicit examples for four point functions, where external particles are scalars which exchange
higher spin fields i.e., consider an analogy of Veneziano amplitude.
 We consider charged scalars since this allows a coupling to odd spins as well.
 A coupling of an irreducible higher spin mode with two scalars is described by the cubic vertex  \cite{Berends:1985xx}
\begin{equation} \label{Berends}
{\cal L}_{int}^{00s}=   \kappa^{1-h}  N_{h} \ {\Psi_h^{\mu_1\dots \mu_h}
J^{1;2}_{h; \mu_1\dots \mu_h}\over h!} + \ h.c.\,
\end{equation}
where
\begin{equation}\label{Jp}
J^{1;2}_{h;\mu_1\dots \mu_h}=\sum_{r=0}^{h} \ \bn{h}{r} \ (-1)^r \
(\partial^{\mu_1} \dots
\partial^{\mu_r}
\phi_1)\ (\partial^{\mu_{r+1}} \dots
\partial^{\mu_{h}}\phi_2)\,
\end{equation}
The consideration of only cubic interactions does not determine the constants  $N_h$.
 We examine two different possibilities
 $N_h= h!$ - ``field theory'' coupling, when  the coupling does not depend on spin and
$N_h= \sqrt{h!}$ - ``string theory'' coupling. The form of the latter coupling can be deduced from the vertices considered in
the previous subsection and they correspond to the fields belonging to leading Regge trajectories. One can also consider reducible higher spin modes
and decompose them into irreducible ones following \cite{Fotopoulos:2009iw}.

The corresponding four point functions can be computed in two different ways:
either using the standard technique of Feynman diagrams,
or using spinorial representations for helicities described earlier in this Section.
In particular,  one can show that ${\cal
M}_4^{(1,2)}(0)$ is given by the expression
\cite{Benincasa:2007xk}
\begin{eqnarray}\label{M12}
{\cal M}_4^{(1,2)}(0) = && \sum_{h > {\rm
max}(-(h_1+h_4),(h_2+h_3))} \big(
\kappa^A_{1-h_1-h_4-h}\kappa^H_{1+h_2+h_3-h}
\frac{(-P_{3,4}^2)^h}{P_{1,4}^2}\left(\frac{[1,4][3,4]}{[1,3]}\right)^{h_4}
 \nonumber \\
&& \left(\frac{[1,3][1,4]}{[3,4]}\right)^{h_1}\left(\frac{\langle
3,4\rangle}{\langle 2,3\rangle\langle
2,4\rangle}\right)^{h_2}\left(\frac{\langle 2,4\rangle}{\langle
2,3\rangle\langle 3,4\rangle}\right)^{h_3}\big)  \\ \nonumber
&&+ \sum_{h
> {\rm
max}(-(h_1+h_3),(h_2+h_4))}\!\!\!\!\!\!\!\!(4\leftrightarrow
3).\nonumber
\end{eqnarray}
where $P_{i,j}= p_i +p_j$ and coupling constant $k_H$ ($k_A$) is
required to give the right dimension to the part of the amplitude
which is holomorphic (antiholomorphic) in the spinor variables
$\lambda^{(i)}$ ($\tilde{\lambda}^{(i)}$). The subscript of the coupling
constants denotes their mass dimension.
The amplitude ${\cal M}_4^{(1,2)}(0)$ can be obtained from (\ref{M12})
by simply exchanging labels $2$ and $4$.

Now let us apply the formula (\ref{M12}) to the tree--level scattering process
for two scalars with the same charge
$\phi(p_1) \phi(p_2) \rightarrow \phi(-p_3) \phi(-p_4)$.
Considering first the ``field theory'' coupling i.e., taking $N_h= h!$
in (\ref{Berends}) one obtains for coupling constants $k_H=k_A= \kappa^{1-h}$
and therefore
\begin{equation}\label{M12FT}
{\cal M}_4^{(1,2)}(0)= \sum_{h \in  \ \mathbb{Z}} \kappa^{2-2h}
(-P^2_{3,4})^h \left( {1\over P^2_{1,4}} + {1\over P^2_{1,3}}
\right),
\end{equation}
$$
{\cal M}_4^{(1,4)}(0)= \sum_{h \in  \ \mathbb{Z}} \kappa^{2-2h}
(-P^2_{1,4})^h  {1\over P^2_{1,3}} ,
$$
 or in terms of the Madelstam variables
\begin{equation} \label{ftc}
{\cal M}_4^{(1,2)}(0)= \kappa^2 {1 \over 1+ \kappa^{-2} s} \left(
{-s\over t u}\right), \quad
{\cal M}_4^{(1,4)}(0)= \kappa^2 {1 \over 1+ \kappa^{-2} t} \left(
{1\over  u}\right).
\end{equation}
Let us note that in the expression for ${\cal M}_4^{(1,4)}(0)$ the s--channel amplitude
is absent due to the charge conservation.
It is useful to compute the same amplitude using standard Feynman diagram
techniques \cite{Bekaert:2009ud} for this specific choice of coupling constants.
Doing so one obtains in t-channel:
\begin{equation}\label{t4}
{\cal M}^t_4 \sim {\kappa^{2}\over t} \left( {1 \over 1 +
{\kappa^{-2}\over 4} (\sqrt{s} +\sqrt{-u})^2}+ {1 \over 1 +
{\kappa^{-2} \over 4}(\sqrt{s} -\sqrt{-u})^2}-1\right).
\end{equation}
In a similar way one can compute the u-channel amplitude and add it to
(\ref{t4})  in order to compare to ${\cal M}_4^{(1,2)}(0)$
given in (\ref{ftc}).
The amplitudes (\ref{t4}) vanish at the complex infinity after making a BCFW shift
 and taking $z \to \infty$ which means that the theory should
 be constructable.
Obviously  four point amplitudes (\ref{ftc}) do not pass the four particle test
(\ref{BC1}). This means in turn that either the theory must have either a trivial S--matrix  or there is
a pathology in its definition.

As one can see from
(\ref{t4})  there is a pole which depends on kinematic variables i.e., there should be some extended object
in the theory. The BCFW computation tells us the same thing
through its failure of crossing symmetry under BCFW deformations,
which suggests that at some finite distance on the complex
kinematic variables plane the massless theory has some ingredient
missing in its definition.  Performing the BCFW shift in
(\ref{t4}) for the
particles $1$ and $2$
  one can easily verify that there are three poles on the
complex z-plane for the t-channel contribution and three for the
u-channel, respectively. The one which comes from the massless
t-pole of (\ref{t4}) is the one whose $Res ({\cal M}(z)/z)$
reproduces the t-channel pole of (\ref{M12FT}). On this pole the
amplitude factorizes into two scalar three point amplitudes which
are, however, dressed with some form factors. The other two poles of
(\ref{t4}) should come from the "extended object" poles which
contribute the form factors of (\ref{ftc}). To get the
u-channel pole of (\ref{M12FT}) we will need to compute the
residues of the u-channel amplitude ${\cal M}_4^u$ in a similar
manner.

Now one can repeat the same arguments for the string theory coupling.
In particular, one can see that the crossing symmetry condition (\ref{BC1})
fails again since four point functions computed with the help of (\ref{M12})
\begin{equation}
 {\cal M}_4^{(1,2)}(0)=e^{-\kappa^{-2}\,s}\left( \frac{-s}{tu} \right),
\quad {\cal M}_4^{(1,4)}(0)=e^{-\kappa^{-2}\,t}\left( \frac{1}{u} \right)
\end{equation}
are not equal to each other. A
similar conclusion can be made either by directly evaluating four point functions
using Feynman diagram techniques or constructing quartic vertices using
the BRST approach \cite{Fotopoulos:2010ay}. In the latter approach one introduces four Hilbert
spaces and has therefore BRST charges $Q_M$, $M=1,2,3,4$.
 Considering for simplicity a part of the cubic vertex (\ref{TOTAL})
which contains only $\Delta_1$
one arrives to the equation for the quartic vertex
\begin{equation} \label{QEQ}
(Q_1+ Q_2 + Q_3 + Q_4) |V_4 \rangle =  18 (Z_{a'1} c^{+ a'} + Z_{a 1}c^{+a}) e^{Y_{1i'}Y_{1i} p^{i'}_\mu p^{i}_\mu} e^M c_0^2 c_0^3c_0^{2'} c_0^{3'} |0 \rangle_{232'3'}
\end{equation}
where
\begin{equation}\label{M}
M= Y_{a i} \alpha^{+ a}_\mu p_\mu^i + Y_{a' i'} \alpha^{+ a'}_\mu p_\mu^{i'} + Z_{ab}c^{+a}b_0^{b} +  Z_{a'b'}c^{+a'}b_0^{b'}
\end{equation}
with $i/i'=1,2,3/1,2',3'$, and  $a,b/a'b'=2,3/2'3'$. The equation (\ref{M})  can be solved to give
\begin{equation}\label{V4}
|V_4 \rangle = F(p) e^M c_0^2 c_0^3c_0^{2'} c_0^{3'} |0 \rangle_{232'3'}, \quad
F(p)= 18\frac{e^{Y_{1i'}Y_{1i} p^{i'}_\mu p^{i}_\mu}}{(p^2_\mu+ p^3_\mu)(p^{2'}_\mu+ p^{3'}_\mu )}.
\end{equation}
Obviously the full solution is given by acting with the above
given vertex on all non-cyclic permutations of the external states
${\cal L}_4\sim \langle 1,2,3,4| V_4\rangle_s + \langle 1,3,2,4|
V_4\rangle_u + \langle 1,4,2,3| V_4\rangle_t $ where
each contribution has subscript indicating the massless pole on
the corresponding kinematic variable which comes from the
definition of $F(p)$ in (\ref{V4}).

 Computing the four point function with the Feynman rules and using the interaction vertices given above
\begin{equation}\label{CEstregge}
{\cal M}_4^t \sim -\frac{\kappa^{2}}{t}\
\exp\!\left(-\kappa^{-2}\,s\right) \ .
\end{equation}
On the other  side computing 
the four point function for tachyons using the Pomeron vertex operator in the bosonic string and  taking the naive
$\alpha^\prime \rightarrow \infty$ limit gives
\begin{equation}\label{HSpomeron}
{\cal M}_4^t \sim {1\over \sqrt{t}}\left( {t\over
s}\right)^{\kappa^{-2} t-1} e^{-\kappa^{-2} t}
\end{equation}
where $\kappa$ is the critical string length we have assumed in order
for our formulas to make sense. This should be compared to
(\ref{CEstregge}) where we see the marked difference between the two
expressions: for the region $s\gg t$ where the two expressions
apply one goes as $e^{-\kappa^{-2} s}$ and the other one has the
typical Pomeron behavior $s^{\kappa^{-2} t-1}$. Our discussion
suggests that massless current exchanges alone cannot lead to
the above behavior  and one can assume that the expression in
(\ref{HSpomeron}) is the appropriate asymptotic behavior of the
tensionless limit of the four point string amplitude for BCFW
deformations $s\sim-u\sim q z$.

Therefore the analysis of the examples above suggests that  interacting higher spin theories
in flat background with only point particle states although not
improbable, might not be the ones related to the high energy limit
of string theory. Although one can add perturbatively quartic and
higher vertices in the Lagrangian, consistently with gauge
invariance at each order, a non-trivial S--matrix, its analyticity
properties and BCFW constructibility seem to imply that: one
should consider higher spin theories in a larger frame which
includes extended and/or non-local objects in their spectrum.\\

\noindent {\bf Acknowledgments.} I am  grateful to
I. Bandos, X. Bekaert, A. Fotopoulos and D. Sorokin
for numerous discussions and collaboration
on the topics presented in the review.
I would like especially thank A. Fotopoulos for reading the manuscript and for his comments.
 I am grateful to my family
who turned my stay in Auckland, New Zealand into a wonderful  experience.
\\

\renewcommand{\thesection}{A}

\setcounter{equation}{0}

\renewcommand{\theequation}{A.\arabic{equation}}

\end{document}